\documentclass{jnst} 
\usepackage[tablesfirst,notablist]{endfloat} 
\usepackage{braket}
\usepackage{xcolor}

\title{Physics-based method for generating probability table using random-matrix approach}
\author{Kazuki Fujio$^{1}$\thanks{Corresponding author. Email: kazuki.fujio@lanl.gov}, 
Toshihiko Kawano$^{1}$, 
Amy Elizabeth Lovell$^{1}$,
Denise Neudecker$^{1}$, and
Noah Anthony Wy Walton$^{1}$}
\inst{$^{1}$Los Alamos National Laboratory, Los Alamos, New Mexico 87545, USA}
\abst{We develop a new method for generating probability tables based on a solid theoretical foundation.
The fluctuating cross sections are calculated using the GOE-$S$-matrix model, in which the Gaussian Orthogonal Ensemble (GOE) is incorporated into the calculation of the scattering ($S$) matrix. 
The calculated cross sections are then converted into the probability tables in the same manner as in NJOY.
Using $^{238}$U and $^{239}$Pu as target nuclei, we determine the optimal model parameters based on the convergence behavior of the average cross sections.
The statistical uncertainty of the probability tables is examined as a function of the number of ladders.
We demonstrate that the probability tables calculated at 0 K are qualitatively comparable with those calculated using the conventional single-level Breit–Wigner formalism, albeit we observe some local differences due to requisite unitality for the $S$ matrix.
}
\kword{Nuclear reaction; Unresolved Resonance Region; Probability table; Random-matrix approach; Gaussian Orthogonal Ensemble}

\begin{document}
\maketitle
\newpage

\section{Introduction}
The neutron-induced reaction cross section is characterized by the ratio of the average resonance width $\braket{\Gamma}$ to the average resonance spacing $D$.
In the low-energy region, because the ratio $\braket{\Gamma}/D$ is small, sharp and isolated resonances are observed; this region is referred to as the resolved resonance region (RRR).
As incident neutron energy increases, $\braket{\Gamma}/D$ becomes larger, and individual resonances begin to overlap, making them experimantally indistinguishable from one another.
As a result, the fluctuation in the cross section gradually diminishes, and the energy dependence of the cross section eventually becomes smooth which is often called the ``fast'' region.
The transition region from the RRR to the smooth cross section region is known as the unresolved resonance region (URR), which typically occurs in the keV energy range for actinide nuclei.

Since the statistical nature of resonances changes in the URR, both the average and fluctuating cross sections influence neutron absorption and reaction rates in reactor analyses.
The self-shielding in the URR is particularly important, as it has been shown to impact the calculations of reactivity~\cite{Weinman1998,Mosteller1999} and neutron flux~\cite{Carter1998}.
The self-shielding effect can be evaluated using Monte Carlo particle transport codes such as Geant4~\cite{Allison2016,Zmeskal2025}, MCNP~\cite{Rising2025}, MVP~\cite{Nagaya2025}, OpenMC~\cite{Romano2015}, and Serpent~\cite{Leppanen2025}, whereas the corresponding continuous energy nuclear data are prepared in the form of probability tables~\cite{Otter1971,Levitt1972,Cullen1974,Blomquist1980} by nuclear data processing codes such as CALENDF~\cite{Sublet2011}, FRENDY~\cite{Tada2024}, and NJOY~\cite{Macfarlane2017}.
The probability table represents the fluctuations of cross sections in the URR in a statistical manner.
For each incident neutron energy, it provides the probability of the total cross section lying within a specific range, along with the corresponding average cross sections for the total, elastic scattering, radiative capture, and fission reactions.
In the URR, the true details of individual resonances are unknown, and only their average properties and probability density functions are provided.
As a result, it is often necessary to generate multiple sets of randomly sampled resonances (ladders) to properly account for these statistical fluctuations.
However, the method to generate ladders is computationally demanding.
Therefore, studies on improving its efficiency have been conducted, which include, e.g., investigations into the optimal number of ladders~\cite{Tada2023} and approaches that parameterize the cross section distribution directly using the normal inverse Gaussian (NIG) distribution without generating ladders~\cite{Ridley2024}.

Looking at the overall cross sections of neutron-induced reactions, the theoretical frameworks for describing cross sections in the RRR and the fast energy region have already been well established by the $R$-matrix theory~\cite{Lane1958} and the Hauser-Feshbach theory~\cite{Hauser1952}, respectively.
However, cross section fluctuations in the URR have been calculated under simplified assumptions with resonance formulas~\cite{Koyumdjieva2010,Koyumdjieva2011,Holcomb2017}.
The single-level Breit-Wigner (SLBW) formula~\cite{Breit1936} or the $R$ matrix with Reich-Moore approximation~\cite{Reich1958} is employed with the Wigner distribution~\cite{Wigner1951} for resonance level spacing and the $\chi$-squared distribution with $N_{\chi}$ degrees of freedom for the decay width.
These approaches have certain deficiencies: the SLBW formula does not properly account for interference between resonances, both approaches need to know the statistical distributions of resonances, and suffer from broken unitarity.
Therefore, a solid theoretical foundation is required for describing cross sections in the URR.

In recent years, there has been active research into the statistical properties of compound nuclei using concepts from random-matrix theory~\cite{Weidenmueller2009,Mitchell2010,Bertsch2018,Brown2018,Fanto2018,Nobre2023,Bertsch2024,Weidenmuller2024}.
Wigner proposed that the statistical properties of random matrices resemble those of resonance level spacings in nuclei~\cite{Wigner1951}.
As an example, the eigenvalue-spacing distribution of the Gaussian Orthogonal Ensemble (GOE)---a set of real symmetric matrices with elements randomly sampled from a Gaussian distribution---shows remarkable agreement with the experimentally observed distribution of nuclear resonance spacing.
Verbaarschot, Weidenm\"uller, and Zirnbauer directly incorporated the random-matrix theory into the scattering matrix formalism by replacing the Hamiltonian with a GOE, and analytically evaluated the average $S$ matrix through a three-fold integration~\cite{Verbaarschot1985}.
Their model, which employs the GOE to calculate the $S$ matrix, is hereafter referred to as the GOE-$S$-matrix model, to distinguish it from the GOE itself.
Kawano, Talou, and Weidenm\"uller further developed the GOE-$S$-matrix model to allow calculations of the average $S$ matrix using Monte Carlo techniques and confirmed that the obtained results are equivalent to those derived from the three-fold integration~\cite{Kawano2015}.
In our previous study, we extended the model to account for actual cross sections by introducing a phase shift~\cite{Fujio2025}.

The objectives of this study are to develop a method for generating the probability table using the GOE-$S$-matrix model.
The GOE-$S$-matrix model includes several parameters required to calculate cross sections.
Using $^{238}$U and $^{239}$Pu as target nuclei, we determine the optimal model parameters based on the convergence behavior of the average cross sections.
Section~\ref{sec:Method} introduces the employed models. 
In this study, we convert the calculated cross sections into the probability tables in the same manner as in NJOY, as described in Sec.~\ref{sec:Method}. 
In Sec.~\ref{sec:Results}, we examine the convergence of the probability table by evaluating the root mean square percentage error (RMSPE), and also compare the obtained tables with those calculated by the SLBW formula at 0 K, in order to make a purely model-based comparison by eliminating temperature effects.
The concluding remarks are summarized in Sec.~\ref{sec:Conclusion}.

\section{Models and calculation method}
\label{sec:Method}
\subsection{GOE-$S$-matrix model}
\label{sec:GOE-S-matrix}
The GOE-$S$-matrix model requires several input parameters.
These include the number of channels $m$ and levels $n$, which determine the size of the model space, as well as the transmission coefficients for each channel, and the channel radius that remaps the GOE-$S$-matrix model onto the actual experimentally observed cross sections. 
The $S$ matrix corresponding to a reaction from channel $a$ to $b$ is expressed as:
\begin{eqnarray}
\label{eq:Smat}
    S^{\rm (GOE)}_{ab}&=&e^{-i\left(\phi_a+\phi_b\right)}\left\{\delta_{ab}-2i\pi\sum_{\mu\nu}W_{a\mu}\left(D^{-1}\right)_{\mu\nu}W_{\nu b}\right\},
\end{eqnarray}
where
\begin{eqnarray}
\label{eq:denom}
    D_{\mu\nu}=E_{\lambda}\delta_{\mu\nu}-H^{\rm (GOE)}_{\mu\nu}+i\pi\sum_cW_{\mu c}W_{c\nu}.
\end{eqnarray}
The GOE Hamiltonian $H^{\rm (GOE)}_{\mu\nu}$ is a time-reversal invariant real symmetric matrix of dimension $n\times n$, with elements randomly sampled from a Gaussian distribution.
The mean value is 0, and the second moments of the off-diagonal elements are $\lambda^2/n$, while those of the diagonal elements are $2\lambda^2/n$, where $\lambda$ is a scaling factor that influences the level spacing; i.e., 
\begin{eqnarray}
    \overline{H^{\rm (GOE)}_{\mu\nu}H^{\rm (GOE)}_{\rho\sigma}}=
    \frac{\lambda^2}{n}(\delta_{\mu\rho}\delta_{\nu\sigma}+\delta_{\mu\sigma}\delta_{\nu\rho}),
\end{eqnarray}
where all $\delta$'s denote the Kronecker delta, and the indices $\mu$, $\nu$, $\rho$, and $\sigma$ run from 1 to $n$.
Within the GOE-$S$-matrix framework, energy $E_{\lambda}$ is defined in the eigenchannel space and expressed in units of $\lambda$. 
The energy $E_{\lambda}$ is mapped onto the incident neutron energy by being orthogonal to it, and $E_{\lambda}=0$ (at the center of the semicircle) is taken as the corresponding actual incident neutron energy~\cite{Fujio2025}.
The eigenvalues of $H^{\rm (GOE)}_{\mu\nu}$ are distributed according to the Wigner's semicircle law over the range $E_{\lambda}=-2\lambda$ to $2\lambda$.
The hard-sphere phase shift $\phi_i=k_ir_i$, where $k_i$ is the wave number for channel $i$ $(i=a,b)$ and $r_i$ is the channel radius, is introduced by analogy with the collision-matrix formulation.
While the radius $r_a$ is determined to reproduce the experimental total cross section, $r_b$ drops out when calculating non-elastic cross sections via $|S^{\rm (GOE)}_{ab}|^2$, where $a\ne b$.
The input transmission coefficient $t_a$ for channel $a$ is used to construct the $n\times m$ coupling strength matrix $G$:
\begin{eqnarray}
    w_{a}&=&\sqrt{\frac{x_a}{\pi}}, \\
    x_{a}&=&\frac{2}{t_a}(1-\sqrt{1-t_a})-1,
\end{eqnarray}
where $w_a$ denotes the coupling vector elements of $G$.
The $n\times m$ matrix $W$ is then calculated from $G$ and an orthogonal matrix $\mathcal{O}$, whose columns are the eigenvectors of $G^{\top}G$:
\begin{eqnarray}
    W=G\mathcal{O}.
\end{eqnarray}
The matrix elements of $W$ are used in Eqs.~(\ref{eq:Smat}) and (\ref{eq:denom}).

The statistical properties of the GOE-$S$-matrix model are characterized by the channel transmission coefficients, without requiring any assumptions about the statistical distributions of level spacings or decay widths.
The generalized transmission coefficients used in this work are calculated with CoH$_3$~\cite{Kawano2021}, employing the optical model potential parameters for the elastic scattering channel proposed by Soukhovitskii, \textit{et al}~\cite{Soukhovitskii2004}.
For neutron capture channel, we employ the giant dipole resonance model.
The Kopecky-Uhl model~\cite{Kopecky1990} is employed for E1 transitions, while the Brink-Axel model~\cite{Brink1957,Axel1962} is applied to higher multi-polarities, from M1 to E3 transitions, including the M1 scissors mode~\cite{Mumpower2017}.
The transmission coefficients for fission channels are calculated by the Hill-Wheeler model~\cite{Hill1953}.
The number of inelastic scattering channels is determined by possible number of neutron partial waves and the corresponding excited levels.
We consider the first $2^+$ excited state at 44.916 keV for $^{238}$U and the first $3/2^+$ excited state at 7.861 keV for $^{239}$Pu in this work~\cite{Capote2009}.
The level density is calculated using the Gilbert-Cameron formula~\cite{Gilbert1965}, with a level density parameter~\cite{Kawano2006} in which the shell correction and pairing energy are determined based on the KTUY05 mass formula~\cite{Koura2005}.
To reproduce the average level spacings of 20.3 eV for $^{239}$U and 2.2 eV for $^{240}$Pu, the level density parameter is respectively adjusted~\cite{Capote2009}.
Discrete level data have been adopted from the RIPL-3~\cite{Capote2009}.
Both discrete and continuum states contribute to neutron capture and fission reactions.
Therefore, the numbers of these reaction channels is not strictly defined due to the binning of the continuum state.
In our previous study, the neutron capture channel was divided into 10 channels to avoid discrepancies in the average cross sections between the GOE-$S$-matrix model and the Hauser-Feshbach theory.
Following the same procedure, we divide the neutron capture and fission channels into 10 channels each.
The number of inelastic channels is determined according to the corresponding excitation levels and the set of allowed neutron partial waves.

The cross section for a reaction from channel $a$ to $b$ is defined by
\begin{eqnarray}
\label{eq:cs}
    \sigma^{\rm (GOE)}_{ab}=\frac{\pi}{k_a^2}g_J\left|\delta_{ab}-S^{\rm (GOE)}_{ab}\right|^2,
\end{eqnarray}
where $g_J$ is the spin factor.

\subsection{Single-level Breit-Wigner formula}
To illustrate the difference of our approach from the conventional probability table method, we also employ the Single-level Breit-Wigner (SLBW) formalism to calculate fluctuating cross sections, using average resonance parameters provided in evaluated nuclear data libraries such as JENDL-5~\cite{Iwamoto2023}, ENDF/B-VIII.0~\cite{Brown2018_endf}, and JEFF-3.3~\cite{Plompen2020}.
For each possible orbital angular momentum $l$ and total angular momentum $J$, the parameters include the average level spacing $D_{l,J}$ and the average widths for the reduced neutron $\braket{\Gamma_{n}^{l,J}}$, radiative neutron capture $\braket{\Gamma_{\gamma}^{l,J}}$, fission $\braket{\Gamma_{f}^{l,J}}$, and other competing reaction channels, such as inelastic scattering $\braket{\Gamma_{n'}^{l,J}}$.
The fluctuating cross sections are calculated using these average parameters and the corresponding statistical distributions.
For the statistical sampling, the level spacing and decay widths are calculated using random numbers drawn from the Wigner distribution $P_{\rm W}$ and a $\chi$-squared distribution with $N_{\chi}$ degrees of freedom, respectively. 
Typically, $N_{\chi}$ is set to 1, which corresponds to the Porter-Thomas distribution $P_{\rm PT}$.
The distributions are given by
\begin{eqnarray}
\label{eq:W}
    P_{\rm W}(x_{\rm W}) &=& \frac{\pi}{2}x_{\rm W}\exp{\left(-\frac{\pi x_{\rm W}^2}{4}\right)}, \\
\label{eq:PT}
    P_{\rm PT}(x_{\rm PT})&=&\frac{1}{\sqrt{2\pi x_{\rm PT}}}\exp{\left(-\frac{x_{\rm PT}}{2}\right)},
\end{eqnarray}
where $x_{\rm W}$ and $x_{\rm PT}$ denote the level spacing and width, each normalized by its corresponding average value.

The elastic scattering cross section ${\sigma_{aa}}$ and a reaction cross section from channel $a$ to $b$, ${\sigma_{ab}}$, are calculated by summing over $l$ and $J$~\cite{ENDFmanual}
\begin{eqnarray}
\label{eq:SLBW_res}
    \sigma_{aa}(E)&=&\sum_l\left\{(2l+1)\frac{4\pi}{k_a^2}\sin^2\phi_l\right. \nonumber \\
    &&+\left.\frac{\pi}{k_a^2}\sum_Jg_J\sum_{r}\frac{\Gamma^{l,J\,2}_{n,r}-2\Gamma^{l,J}_{n,r}\Gamma^{l,J}_{tot,r}\sin^2\phi_l+2(E-E'_r)\Gamma^{l,J}_{n,r}\sin(2\phi_l)}{(E-E_r')^2+\frac{1}{4}\Gamma^{l,J\,2}_{tot,r}}\right\},  \\
    \sigma_{ab}(E)&=&\frac{\pi}{k_a^2}\sum_{l,J}g_J\sum_r\frac{\Gamma^{l,J}_{n,r}\Gamma^{l,J}_{b,r}}{(E-E'_r)^2+\frac{1}{4}\Gamma^{l,J\,2}_{tot,r}},
\end{eqnarray}
where $E_r'$ denotes the energy of the $r$-th resonance, calculated using $D_{l,J}$ and a random number sampled from $P_{\rm W}$.
Since the neutron capture and fission channels involve an enormous number of final states due to the high level density of the compound nucleus, the Porter-Thomas distribution is not applied to the widths of these channels, i.e., $\Gamma^{l,J}_{\gamma,r}=\braket{\Gamma_{\gamma}^{l,J}}$ and $\Gamma^{l,J}_{f,r}=\braket{\Gamma_{f}^{l,J}}$.
The Porter-Thomas distribution is applied only to the widths of the elastic and inelastic scattering channels.
The width $\Gamma^{l,J}_{tot,r}$ represents the total width at $r$-th resonance, defined as the sum of the partial widths:
\begin{eqnarray}
\label{eq:tot_width}
    \Gamma^{l,J}_{tot,r}=\Gamma^{l,J}_{n,r}+\Gamma^{l,J}_{\gamma,r}+\Gamma^{l,J}_{f,r}+\Gamma^{l,J}_{{n'},r}.
\end{eqnarray}
The hard-sphere phase shift $\phi_l$ is determined from the channel radius specified in the evaluated data.

\subsection{Probability table}
\begin{figure}
    \centering
    \includegraphics[width=\linewidth]{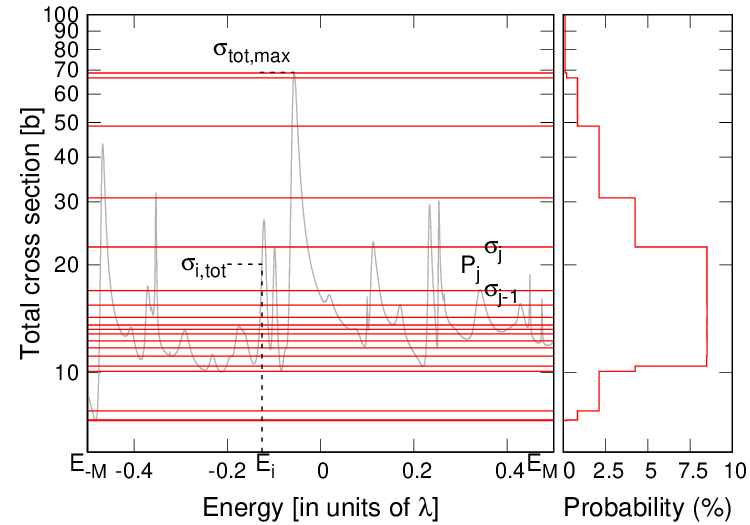}
    \caption{Schematic illustration of the cross section (left) and the probability table generated from the corresponding cross section (right)}
    \label{fig:schematic_ptable}
\end{figure}

While conventional SLBW approaches assume statistical distributions for resonances, the GOE-$S$-matrix model does not require such assumptions, allowing the calculation of probability tables in a more physically grounded manner.
In this model, resonances are distributed in the eigenchannel space, centered at $E_{\lambda}$.
Note that, since a probability table can be calculated for each value of the incident neutron energy, the physical quantities consisting of the probability table depend on the incident energy.
Here, we consider the energy range $E_{-M}$ to $E_{M}$, where the indices $-M$ and $M$ denote the minimum and maximum energy grids in the eigenchannel space.
Figure~\ref{fig:schematic_ptable} shows a schematic illustration of the cross section (left) and the probability table generated from the corresponding cross section (right).
In this figure, $E_{-M}$ and $E_{M}$ are set to $-0.5$ and 0.5, respectively.
The boundary cross section $\sigma_j(E_{\rm inc})$, which defines the upper bound of the $j$-th total cross section bin and corresponds to the horizontal lines shown in Fig.~\ref{fig:schematic_ptable}, is determined by
\begin{eqnarray}
    \sigma_0(E_{\rm inc})&=&0, \nonumber \\
    \sigma_j(E_{\rm inc})&>&\sigma_{j-1}(E_{\rm inc}), \nonumber \\
    \sigma_N(E_{\rm inc})&>&\sigma_{tot,{\rm max}}(E_{\rm inc}),
\end{eqnarray}
where $\sigma_{tot,{\rm max}}(E_{\rm inc})$ is the maximum total cross section in the range $E_{-M}$ to $E_{M}$.
The method to determine $\sigma_j$ in this study, as outlined above, is the same as that employed in NJOY, and the number of $\sigma_j$ is set to $N=20$, which is a commonly adopted value.
The probability $P_j(E_{\rm inc})$ is defined as the probability that the total cross section lies between $\sigma_{j}$ and ${\sigma}_{j-1}$:
\begin{eqnarray}
\label{eq:prob}
    P_j(E_{\rm inc})=\frac{\displaystyle \sum_{i=-M}^{M}\delta_{ij}(E_{\rm inc})}{\displaystyle\sum_{i=-M}^{M}1},
\end{eqnarray}
where 
\begin{equation}
\delta_{ij}(E_{\rm inc})=\left\{ \,
    \begin{aligned}
    & 1,\hspace{5mm} \sigma_{j-1}(E_{\rm inc})\leq\sigma_{i,tot}(E_{\rm inc})<\sigma_j(E_{\rm inc}), \\
    & 0,\hspace{5mm} {\rm otherwise},
    \end{aligned}
\right.
\end{equation}
and $\sigma_{i,tot}$ denotes the $i$-th total cross section, which lies between $E_{-M}$ and $E_{M}$.
The cumulative probability below $P_j(E_{\rm inc})$ is obtained by summing up:
\begin{eqnarray}
    P(E_{\rm inc},\sigma\leq\sigma_j)=\sum^{j}_{k=1}P_k(E_{\rm inc}).
\end{eqnarray}
The average cross section $\bar{\sigma}_{j,x}(E_{\rm inc})$ in the $j$-th bin is written by
\begin{eqnarray}
\label{eq:ave_sigma}
    \bar{\sigma}_{j,x}(E_{\rm inc})=\frac{\displaystyle \sum_{i=-M}^{M}\sigma_{i,x}(E_{\rm inc})\delta_{ij}(E_{\rm inc})}{\displaystyle \sum_{i=-M}^{M}\delta_{ij}(E_{\rm inc})},
\end{eqnarray}
where $\sigma_{i,x}$ denotes the $i$-th cross section, and the index $x$ in $\bar{\sigma}_{j,x}$ and $\sigma_{i,x}$ refers to total, elastic scattering, neutron capture, inelastic scattering, or fission.

\section{Results and discussion}
\label{sec:Results}
\subsection{Dependence of average cross sections on model parameters}
\label{sec:dependence}
The determination of the transmission coefficients and $m$ is described in Sec.~\ref{sec:GOE-S-matrix}.
In this section, the parameters, $n$ and the range of $E_{\lambda}$, are determined by examining the convergence behavior of the average cross sections as a function of the number of ladders $L$.

\begin{figure}
    \centering
    \includegraphics[width=\linewidth]{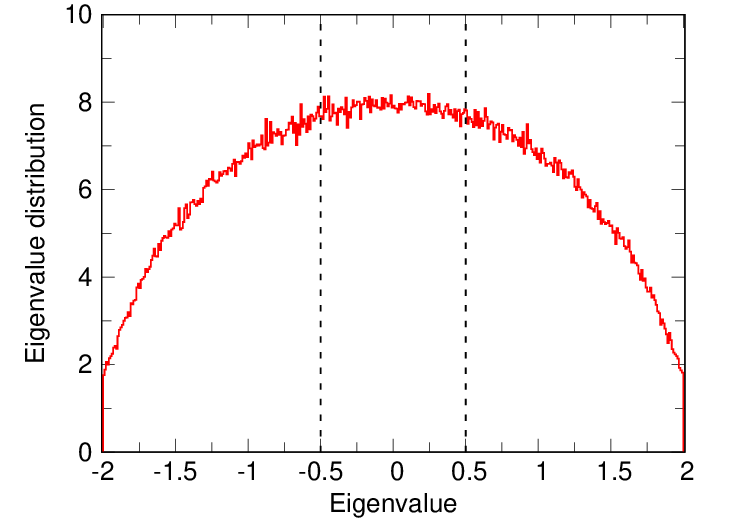}
    \caption{The eigenvalue distribution of 0.1 million realizations of a $10\times10$ GOE.}
    \label{fig:semicircle}
\end{figure}

The eigenvalues of the GOE itself are distributed over the range $E_{\lambda}=-2\lambda$ to $2\lambda$ (hereafter referred to as the full $E_{\lambda}$ range). 
The level density follows the semicircle law, as shown in Fig.~\ref{fig:semicircle}.
The distribution in Fig.~\ref{fig:semicircle} was obtained by 0.1 million realizations of a $10\times10$ dimensional GOE, with the scaling factor $\lambda$ set to 1.
The level density is approximately constant near the center of the semicircle, and the level spacing distribution calculated from resonances within a narrower region around the center more closely resembles the Wigner distribution.
The region enclosed by the vertical dashed lines in Fig.~\ref{fig:semicircle} corresponds to the eigenvalue distribution within the semicircular region from $E_{\lambda}=-\lambda/2$ to $\lambda/2$, which is referred to as the quarter $E_{\lambda}$ range.
Because the distribution in this region is approximately flat, a constant level density is assumed.

\begin{figure}
    \centering
    \includegraphics[width=\linewidth]{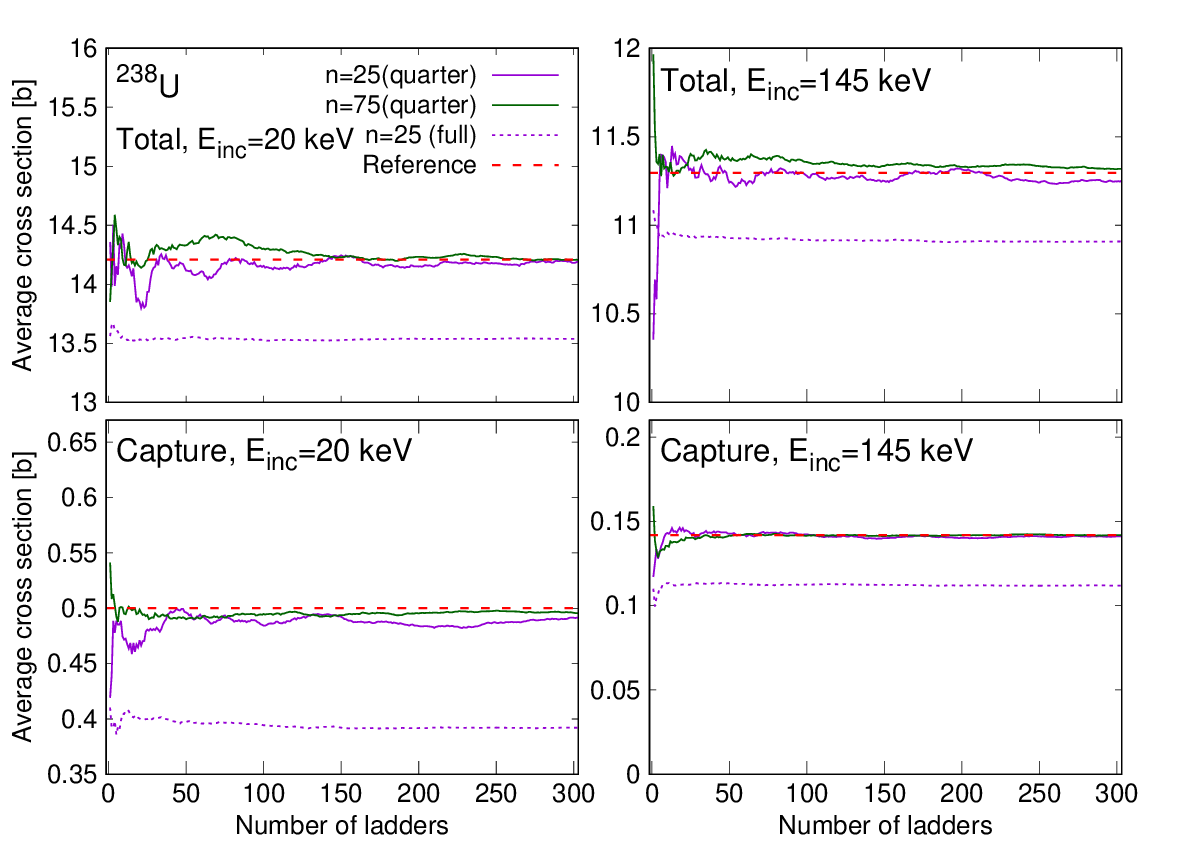}
    \caption{Average total (top) and capture (bottom) cross sections for $^{238}$U as a function of $L$.}
    \label{fig:aveU238}
\end{figure}

\begin{figure}
    \centering
    \includegraphics[width=\linewidth]{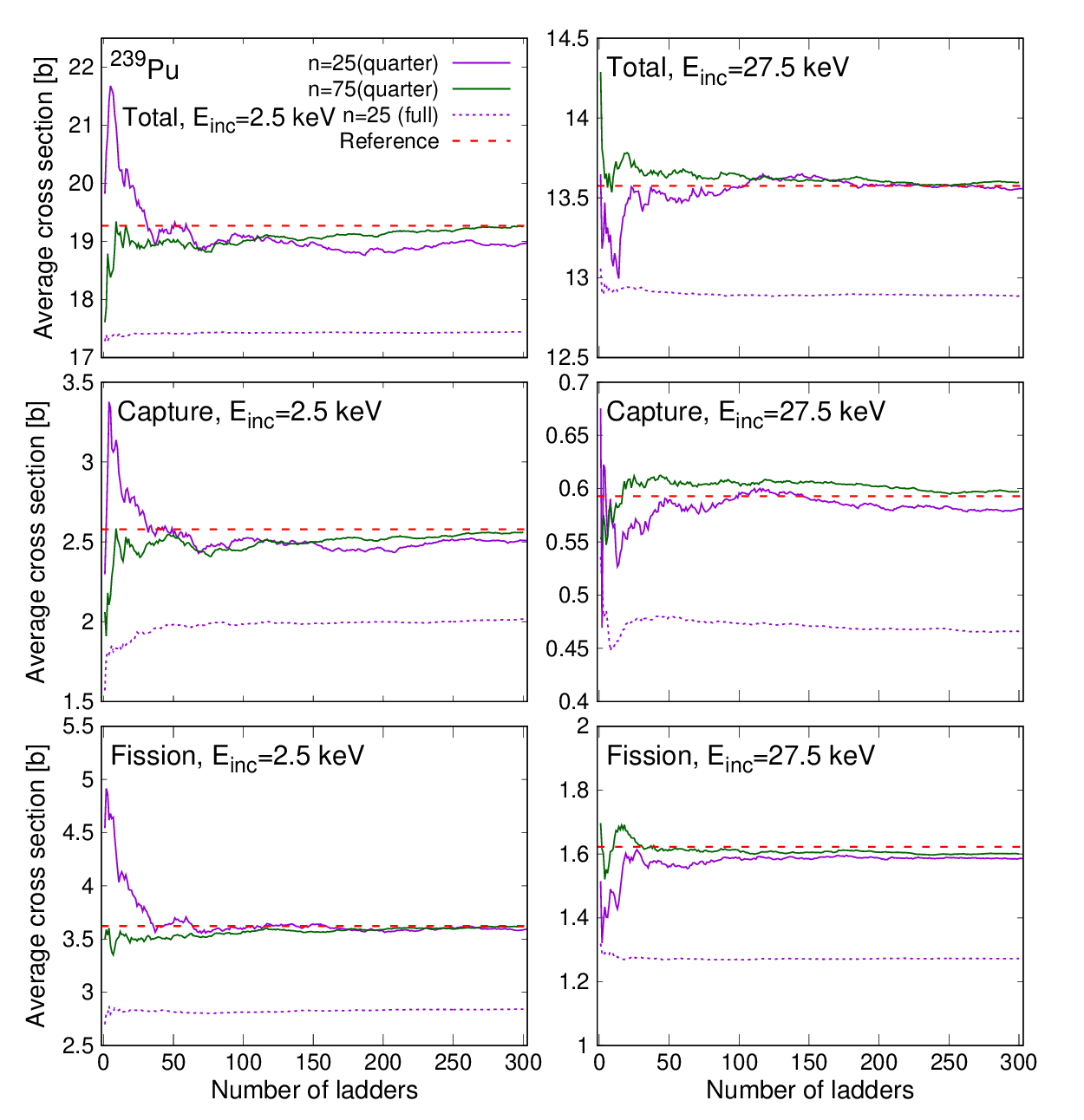}
    \caption{Average total (top), capture (middle), and fission (bottom) cross sections for $^{239}$Pu as a function of $L$.}
    \label{fig:avePu239}
\end{figure}

Figures~\ref{fig:aveU238} and~\ref{fig:avePu239} show the cross sections averaged over these $E_{\lambda}$ range as a function of $L$ for $^{238}$U and $^{239}$Pu.
The horizontal dashed lines correspond to the cross sections averaged over 20 million GOE realizations at $E_{\lambda}=0$ with $n=25$, and are comparable to the evaluated data.
Note that the dependence on $n$ is quite small for this result, since the average is only taken at $E_{\lambda}=0$.
These cross sections shown in the horizontal dashed lines are used as a reference for comparison throughout Sec.~\ref{sec:dependence}.
For clarity, the input $n$ is defined more precisely as the number of resonances within the full $E_{\lambda}$ range.
Therefore, the convergence of the average cross sections in the full $E_{\lambda}$ range  (dotted lines) is relatively faster because of the smaller dispersion in the fluctuating cross sections.
In contrast, since $n$ is not constant in the quarter $E_{\lambda}$ range (solid lines), fluctuating cross sections exhibit a larger dispersion, leading to slower convergence.
The cross sections averaged over the quarter $E_{\lambda}$ range better agree with the reference than those averaged over the full $E_{\lambda}$ range.
For the $^{238}$U case, the deviations obtained with the full $E_{\lambda}$ range are approximately 4\% for the total cross sections and more than 20\% for the neutron capture cross sections. 
Using the quarter $E_{\lambda}$ range reduces the deviations to about 1\% or less for both the total and the capture cross sections.
The same trends are observed in the $^{239}$Pu case. 
The deviations calculated with the full $E_{\lambda}$ range tend to be larger than those obtained with the quarter $E_{\lambda}$ range.
The deviations are confirmed to decrease further as $L$ increases.
We also compare the deviations for $n=25$ and $n=75$ within the quarter $E_{\lambda}$ range. 
As $n$ increases, the deviation converges more rapidly, resulting in higher accuracy; however, the matrix dimension grows and the computational cost increases dramatically.
Hereafter, we use $n=25$ and the quarter $E_{\lambda}$ range as the default settings.

\subsection{Statistical uncertainty evaluated by RMSPE}
The root mean square percentage error (RMSPE) for each reaction is evaluated as a function of $L$ to quantify the statistical uncertainty of the probability table arising from the finite number of GOE realizations used in its construction.
While resonance peaks significantly affect self-shielding in practical transport calculations, the probabilities of the large cross-section bins corresponding to those peaks are relatively small.
To account for this, we calculate the RMSPE of the product of the probability and each average cross section~\cite{Tada2023}:
\begin{eqnarray}
    RMSPE^{(L)}_{x}(\%)=100\times\sqrt{\frac{1}{N}\sum_{j=1}^{N}\left\{\frac{P_{j}^{(L_{\rm ref})}\bar{\sigma}_{j,x}^{(L_{\rm ref})}-P_{j}^{(L)}\bar{\sigma}_{j,x}^{(L)}}{P_{j}^{(L_{\rm ref})}\bar{\sigma}_{j,x}^{(L_{\rm ref})}}\right\}^2}
\end{eqnarray}
where $P^{(L)}_{j}$ and $\bar{\sigma}^{(L)}_{j,x}$ denote the $L$-ladder averages of $P_{j}$ and $\bar{\sigma}_{j,x}$, which are defined in Eqs.~(\ref{eq:prob}) and (\ref{eq:ave_sigma}), respectively.
Here, the reference number of ladders $L_{\rm ref}$ is set to 20000.
Tables~\ref{tab:U238} and \ref{tab:Pu239} present the RMSPE as a function of $L$ for $^{238}$U and $^{239}$Pu.
It was confirmed that the RMSPE decreases for both $^{238}$U and $^{239}$Pu as $L$ increases.
For $^{239}$Pu at 27.5 keV with $L=100$, convergence is slower than for other $L=100$ results. 
This slower convergence is due to the fact that $n$ is not constant in the quarter $E_{\lambda}$ range, as noted in Sec.~\ref{sec:dependence}, which leads to a larger dispersion in the cross section.
The RMSPE falls below 5\% at $L=1000$ and reaches approximately 1\% at $L=10000$.
\begin{table}
\caption{The RMSPE of the product of the probability and the average cross sections in the probability table for $^{238}$U, as a function of $L$.}
\begin{tabular}{ccccccc}
\hline
$E_{\rm inc}$ {[}keV{]} & $L$    & Total   & Elastic & Fission & Capture & Inelastic \\ \hline
20                      & 10    & 11.61\% & 12.30\% & -       & 10.09\% & -         \\
                        & 100   & 3.08\%  & 3.31\%  & -       & 2.87\%  & -         \\
                        & 1000  & 0.84\%  & 0.79\%  & -       & 0.91\%  & -         \\
                        & 10000 & 0.25\%  & 0.23\%  & -       & 0.28\%  & -         \\
\hline   
145                     & 10    & 17.20\% & 16.97\% & -       & 15.76\% & 24.03\%   \\
                        & 100   & 2.95\%  & 3.02\%  & -       & 3.74\%  & 6.44\%    \\
                        & 1000  & 1.69\%  & 1.72\%  & -       & 1.80\%  & 2.21\%    \\
                        & 10000 & 0.65\%  & 0.83\%  & -       & 0.63\%  & 0.54\%    \\
\hline   
\end{tabular}
\label{tab:U238}
\end{table}

\begin{table}
\caption{The RMSPE of the product of the probability and the average cross sections in the probability table for $^{239}$Pu, as a function of $L$.}
\begin{tabular}{ccccccc}
\hline
$E_{\rm inc}$ {[}keV{]} & $L$    & Total   & Elastic & Fission & Capture & Inelastic \\ \hline
2.5                      & 10   & 28.74\% & 29.04\% & 32.02\%     & 24.30\% & -         \\
                        & 100   & 4.32\%  & 4.22\%  &  4.87\%     & 7.60\%  & -         \\
                        & 1000  & 2.63\%  & 2.88\%  &  2.66\%     & 2.39\%  & -         \\
                        & 10000 & 0.28\%  & 0.29\%  &  0.35\%     & 0.21\%  & -         \\
\hline   
27.5                     & 10   & 29.01\% & 28.71\% &  31.66\%    & 38.09\% & 31.49\%   \\
                        & 100   & 23.13\% & 24.74\% &  21.02\%    & 20.46\% & 18.87\%   \\
                        & 1000  &  3.00\% &  3.36\% &   3.15\%    &  2.56\% &  3.31\%   \\
                        & 10000 &  1.14\% &  1.21\% &   1.18\%    &  1.00\% &  0.99\%   \\
\hline   
\end{tabular}
\label{tab:Pu239}
\end{table}

\subsection{Comparison of probability tables}
\begin{figure}
    \centering
    \includegraphics[width=\linewidth]{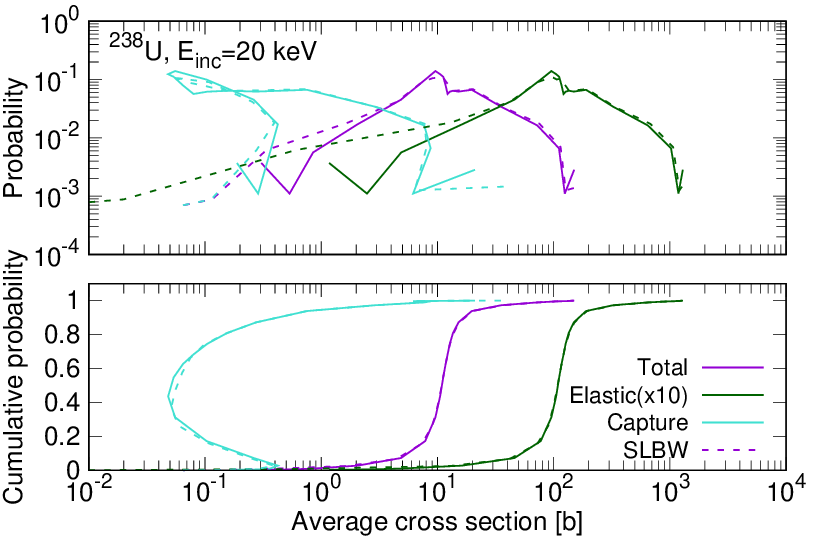}
    \caption{The probability (top) and cumulative probability (bottom) as functions of the average cross sections for $^{238}$U at an incident neutron energy of 20 keV. The results for elastic scattering are scaled by a factor of 10 along the $x$-axis.}
    \label{fig:ptableU238_20keV}
\end{figure}

\begin{figure}
    \centering
    \includegraphics[width=\linewidth]{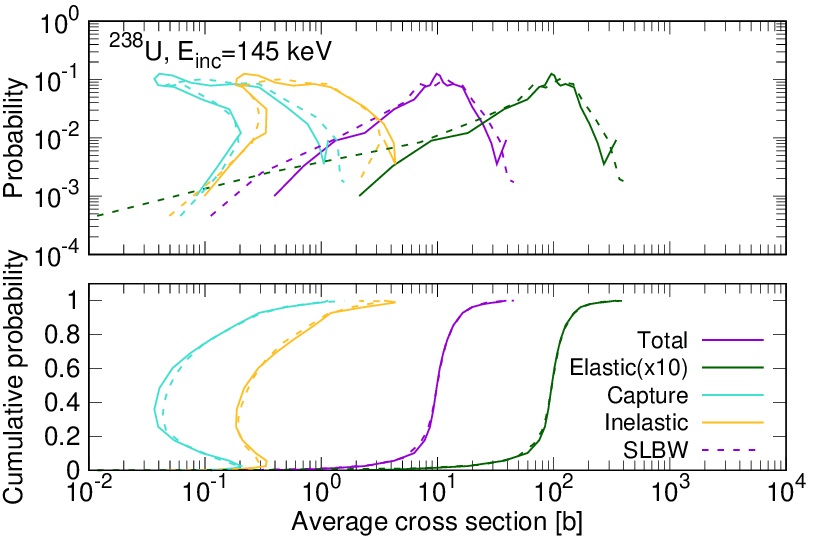}
    \caption{The probability (top) and cumulative probability (bottom) as functions of the average cross sections for $^{238}$U at an incident neutron energy of 145 keV. The results for elastic scattering are scaled by a factor of 10 along the $x$-axis.}
    \label{fig:ptableU238_145keV}
\end{figure}

\begin{figure}
    \centering
    \includegraphics[width=\linewidth]{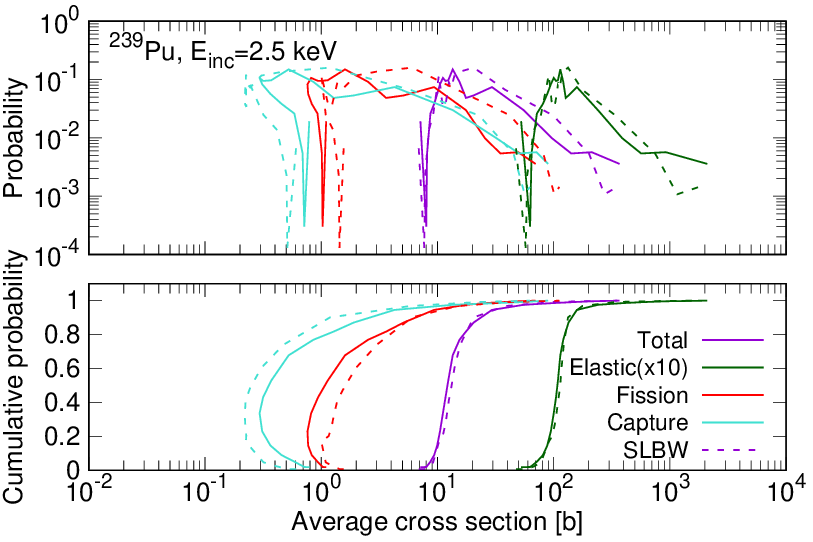}
    \caption{The probability (top) and cumulative probability (bottom) as functions of the average cross sections for $^{239}$Pu at an incident neutron energy of 2.5 keV. The results for elastic scattering are scaled by a factor of 10 along the $x$-axis.}
    \label{fig:ptablePu239_2.5keV}
\end{figure}

\begin{figure}
    \centering
    \includegraphics[width=\linewidth]{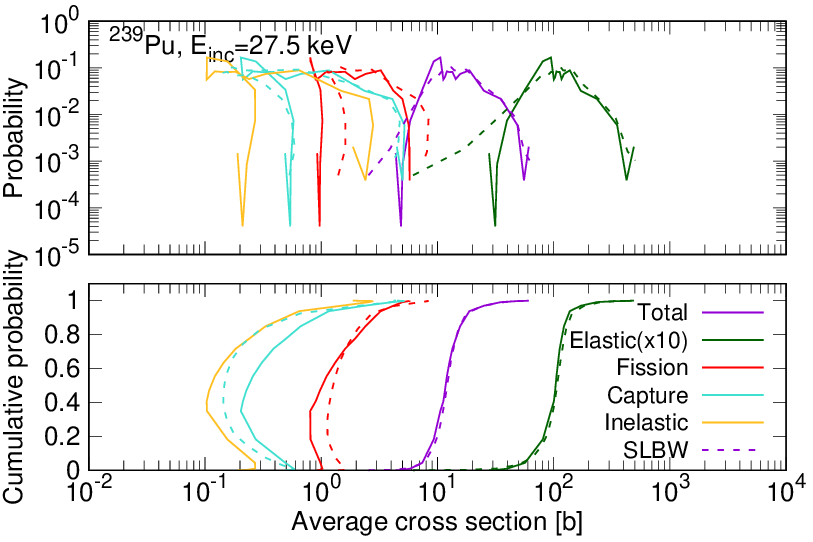}
    \caption{The probability (top) and cumulative probability (bottom) as functions of the average cross sections for $^{239}$Pu at an incident neutron energy of 27.5 keV. The results for elastic scattering are scaled by a factor of 10 along the $x$-axis.}
    \label{fig:ptablePu239_27.5keV}
\end{figure}

The calculated probability and cumulative probability as functions of the average cross section are compared with those derived from the SLBW formula.
Figures~\ref{fig:ptableU238_20keV} and \ref{fig:ptableU238_145keV} illustrate the $^{238}$U results, whereas Figs.~\ref{fig:ptablePu239_2.5keV} and~\ref{fig:ptablePu239_27.5keV} correspond to the $^{239}$Pu results.
In this comparison, results averaged over $L=10000$ are presented.
The calculated probability tables for both $^{238}$U $^{239}$Pu are qualitatively comparable with those calculated using the conventional SLBW formalism.
The cumulative probabilities for total and elastic scattering in the GOE-$S$-matrix model are in good agreement with those obtained from the SLBW formula for both $^{238}$U and $^{239}$Pu.
In the case of $^{238}$U, where the partial width for inelastic scattering is included in the evaluated data, the calculated result agrees with that from the SLBW.
The GOE-$S$-matrix model can treat each inelastic scattering channel in the same manner as in the higher energy region.
However, for $^{239}$Pu, the SLBW approach cannot calculate the inelastic scattering cross section because the evaluated partial widths are often not available.
The differences in the capture and fission results between the GOE-$S$-matrix model and the SLBW approach may arise not only from the differences in the input transmission coefficients and partial widths, but also from the fact that the SLBW treats capture and fission with only a single channel each.

\begin{figure}
    \centering
    \includegraphics[width=\linewidth]{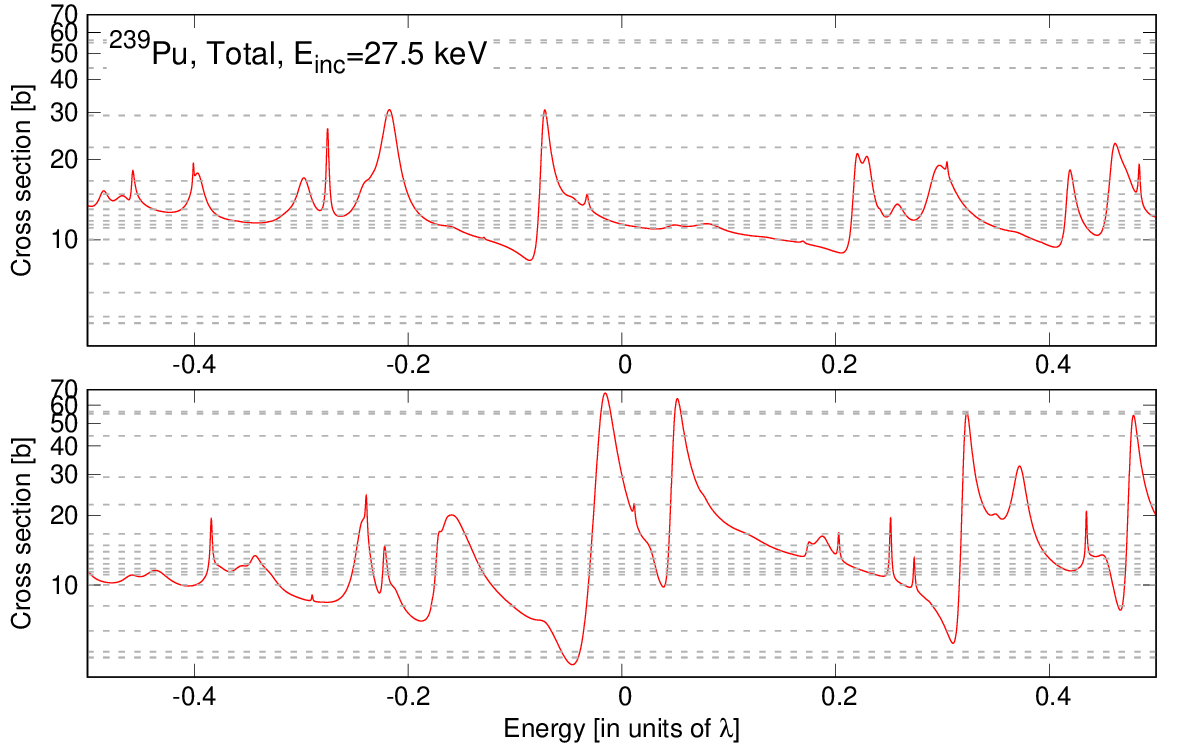}
    \caption{Two examples of total cross sections from the GOE-$S$-matrix model for $^{239}$Pu at an incident neutron energy of 27.5 keV.}
    \label{fig:fluctuating}
\end{figure}

A more pronounced difference is observed in the probability.
The probabilities for the total and elastic scattering obtained with the SLBW approach tend to show small values in the low probability.
This is because the SLBW approach does not guarantee unitarity, which can lead to unphysical negative cross sections and thus to artificially small values in these bins.
Following the same procedure used in NJOY, the negative cross sections are replaced by $1\,\mu{\rm b}$.
In contrast, in the GOE-$S$-matrix model, unitarity is preserved and negative cross sections do not occur.
The probabilities calculated with the GOE-$S$-matrix model also tend to exhibit stronger fluctuations than those obtained with the SLBW approach because of the larger probability spikes at low and high cross section values.
This is because the cross section can vary significantly from event to event, even under identical calculation conditions.
Figure~\ref{fig:fluctuating} shows two examples of the total cross section for $^{239}$Pu at an incident neutron energy of 27.5 keV.
The horizontal dashed lines correspond to the boundaries of cross section bins.
In the top panel of Fig.~\ref{fig:fluctuating}, the resonance amplitudes are relatively weak, and the cross sections do not populate the lowest five bins or the highest three bins.
In contrast, the bottom panel shows strong resonance amplitudes, resulting in larger probabilities in both the lowest and highest bins.
Because the ladder-averaged probability includes such event-to-event variations, it often does not become smooth.
As shown in the probability plots, the smallest and largest bins are particularly sensitive to whether a resonance falls within those bins, which leads to noticeable variations in both the probability and the average cross section.
To reduce fluctuations in the probability distribution, further refinement of the determination of the boundary cross sections will be necessary.

\section{Conclusions}
\label{sec:Conclusion}
We developed a physics-based method for generating the probability table using the GOE-$S$-matrix model.
The GOE-$S$-matrix model employs transmission coefficients, as in the Hauser-Feshbach model, instead of the average resonance parameters in nuclear data libraries.
The probability tables calculated using this method have several advantages compared to those calculated with the conventional SLBW approach.
The GOE-$S$-matrix approach guarantees unitarity and enables and a more realistic treatment of lumped channels, such as capture and fission channels.
Furthermore, because the model can treat inelastic scattering on a channel-by-channel basis, the inelastic scattering cross section for each excited state can be calculated individually.

In the GOE-$S$-matrix model, the resonances are generated over the energy range $E_{\lambda}=-2\lambda$ to $2\lambda$.
By restricting the range to $E_{\lambda}=-\lambda/2$ to $\lambda/2$, where the level density is approximately constant, we found that the ladder-averaged cross sections converge to values comparable to the evaluated data.
Although the convergence became faster and the accuracy improved as the number of levels $n$ increases, the computational time increased dramatically due to the larger matrix dimension; therefore, we concluded that $n=25$ is sufficient.
The statistical uncertainty of the probability table was investigated using the RMSPE.
For each reaction, the RMSPE of the product of the probability and the average cross section was calculated as a function of the number of ladders $L$.
It was found that the RMSPE decreases with increasing $L$, dropping below 5\% at $L=1000$ and approaching 1\% at $L=10000$.
The probability tables calculated using the present model were found to be very similar to those obtained with the conventional SLBW formalism.
The main difference between the GOE-$S$-matrix model and the SLBW lies in the conservation of unitarity.
Also, a noticeable discrepancy was observed in larger probability spikes in the table at low and high cross section values.
We conclude that a reliable method for generating probability tables based on the GOE-$S$-matrix model has been established.
In future work, we will incorporate temperature effects and improve the determination of boundary cross sections.


\section*{Acknowledgement}
Los Alamos National Laboratory is operated by Triad National Security, LLC, for the National Nuclear Security Administration of the US Department of Energy under Contract No. 89233218CNA000001. 
Research reported in this publication was supported by the U.S. Department of Energy LDRD program at Los Alamos National Laboratory (Project No. 20240031DR).

%
\bibliography{reference}
\bibliographystyle{tfnlm}

\end{document}